\documentclass[aps,twocolumn,letterpaper,showpacs]{revtex4}%
\usepackage{amsfonts}
\usepackage{amsmath}
\usepackage{amssymb}
\usepackage[usenames,dvipsnames]{color}
\usepackage{hyperref}
\usepackage{graphicx}%
\providecommand{\U}[1]{\protect\rule{.1in}{.1in}}

\definecolor{orange}{rgb}{1,0.5,0}
\definecolor{pink}{rgb}{1.00,0.75,0.80}
\begin{document}
\title{Universal dynamical decoupling of multiqubit states from environment}
\date{\today}
\author{Liang Jiang,$^{1}$ Adilet Imambekov$^{2}$}
\affiliation{$^{1}$ Institute for Quantum Information, California Institute of Technology,
Pasadena, CA 91125, USA}
\affiliation{$^{2}$ Department of Physics and Astronomy, Rice University, Houston, TX, 77005}

\begin{abstract}
We study the dynamical decoupling of multiqubit states from environment. For
a system of $m$ qubits, the nested Uhrig dynamical decoupling (NUDD) sequence
can efficiently suppress generic decoherence induced by system-environment
interaction to order $N$ using $\left(  N+1\right)  ^{2m}$ pulses. We prove
that the NUDD sequence is universal, i.e., it can restore the coherence of
$m$-qubit quantum system independent of the details of system-environment
interaction. We also construct a general mapping between dynamical decoupling
problems and discrete quantum walks in certain functional spaces.

\end{abstract}

\pacs{03.67.Pp, 03.65.Yz, 76.60.Lz, 82.56.Jn}
\maketitle

Dynamical decoupling (DD) is a powerful tool to protect quantum systems from
decoherence induced by the inevitable system-environment interaction
\cite{Viola99}. The idea of DD is to dynamically control the system (or
environment) evolution to suppress the decoherence caused by interaction. For
example, a static magnetic field with unknown magnitude $B_{z}\sigma_{z}$ can
induce dephasing of a qubit, but such dephasing can be fully eliminated by a
spin flip $\sigma_{x}$ (i.e., Hahn echo) at half way of the evolution
\cite{Hahn50}. In practice, however, the Hahn echo only suppresses the
dephasing to $O\left(  T^{2}\right)  $ for total evolution time $T$, because
the magnetic field may have complicated time-dependence in both magnitude and
orientation due to the evolution of the environment. Furthermore, if the
environment consists of quantum degrees of freedom, it can become entangled
with the system via the interaction. Hence it is a challenging task to design
a \emph{universal} DD scheme that can suppress decoherence to desired order
independent of the details of system-environment interaction.

One particular interesting DD scheme is the concatenated DD (CDD), which has
been shown to be universal for single qubits \cite{Khodjasteh05}. The
limitation, however, is that the pulse number increases exponentially with the
suppression order $N$ (approximately $4^{N}$ pulses to suppress both bit-flip
and dephasing processes to $O\left(  T^{N+1}\right)  $). It is the discovery
of the universality of Uhrig DD (UDD) sequence \cite{Uhrig07,Lee08,Yang08,Pasini10,Uhrig10}
that makes the universal DD practically feasible. In contrast to CDD demanding
exponentially many pulses \cite{Khodjasteh05}, UDD uses only $O\left(
N\right)  $ spin-flip pulses to suppress the dephasing processes to $O\left(
T^{N+1}\right)  $ \cite{Lee08,Yang08}. The discovery of UDD sequence has
inspired many experimental efforts to further improve the coherence over a
wide range of quantum systems, including trapped ions \cite{Biercuk09},
electron spins \cite{deLange10}, defect centers \cite{deLange10,Ryan10},
quantum dots \cite{Bluhm11,Barthel10}, and superconducting qubits
\cite{Bylander11}. However, UDD is restricted to pure dephasing errors of a
single qubit. It is desirable to have an efficient DD scheme (with
$poly\left(  N\right)  $ pulses) to suppress both bit-flip and dephasing
processes for multiple qubits to $O\left(  T^{N+1}\right)  $.

Recently, the quadratic DD (QDD) scheme has been proposed \cite{West10}, which
uses $\left(  N+1\right)  ^{2}$ pulses to suppress both bit-flip and dephasing
errors of single qubits. As a generalization of QDD from $1$-qubit system to
$m$-qubit system, the nested UDD (NUDD) scheme has been proposed
\cite{Mukhtar10a,Mukhtar10b,WangZY11a}, which uses $\left(  N+1\right)  ^{2m}$
pulses to suppress decoherence from the most general interaction between the
$m$-qubit system and environment. Although there are numerical evidences
\cite{West10} and theoretical implications \cite{Pasini10,WangZY11a,WangZY11b} that QDD/NUDD might
be universal, it is still an open question whether QDD/NUDD are universal or
not \cite{Yang10,WangZY11a}.

\begin{figure}[t]
\centering
\includegraphics[width=8.8cm]{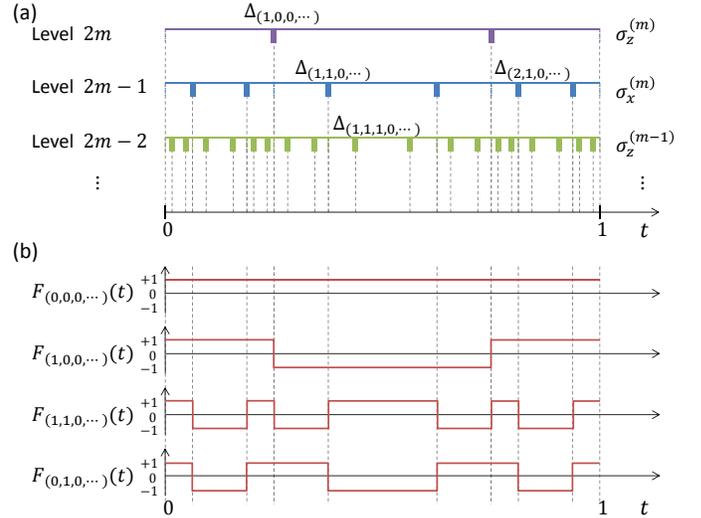}\caption[fig:NUDD]{ (color
online). Nested Uhrig dynamical decoupling (NUDD) scheme with $2m$ nesting
levels and suppression order $N=2$. (a) The timings of NUDD pulses have a
self-similar structure, determined by Eqs.~(\ref{eq:Delta1}) and
(\ref{eq:Delta2}). The set of pulses associated with $r$th level is
$\sigma_{x}^{\left(  j\right)  }$ for $r=2j-1$, and $\sigma_{z}^{\left(
j\right)  }$ for $r=2j$. (b) The time dependent modulation functions
$F_{\alpha}\left(  t\right)  $ of the toggling frame Hamiltonian, and the
corresponding pulses for the $m$th qubit.}%
\label{fig:NUDD}%
\end{figure}

In this Letter, we shall present a rigorous proof that the NUDD scheme with
$2m$ nesting levels and $\left(  N+1\right)  ^{2m}$ pulses is a universal DD
scheme for $m$-qubit system, which suppresses decoherence processes to
$O\left(  T^{N+1}\right)  $ for arbitrary system-environment interaction. We
achieve this by providing a mapping between NUDD and a discrete
\textquotedblleft quantum walk" in $2m$ dimensional space. The rules that
govern this this quantum walk do not depend on $2m$, which allows for an
efficient proof for all nesting levels. Below we first introduce notations and
explain the existing proof for UDD \cite{Yang08,Pasini10,Uhrig10} using the language that can
be naturally generalized for QDD/NUDD.

\paragraph*{UDD.---}

Let us first consider the UDD sequence \cite{Uhrig07} for a qubit-environment
interaction%
\begin{equation}
H\left(  \tau\right)  =\hat{S}_{0}\otimes\hat{B}_{0}\left(  \tau\right)
+\hat{S}_{1}\otimes\hat{B}_{1}\left(  \tau\right)  ,
\end{equation}
where $\hat{S}_{0}=I$ and $\hat{S}_{1}=\sigma_{z}$, and the time-dependent
bath operators are analytic with series expansion $\hat{B}_{\alpha}\left(
t\right)  =\sum_{p=0}^{\infty}\hat{b}_{\alpha,p}t^{p}$ for $\alpha=0,1$. The
UDD sequence uses $N$ $\pi$-pulses (i.e., $\sigma_{x}$ rotations) applied at
times $\tau_{\lambda}=T\Delta_{\lambda}$, where
\begin{equation}
\Delta_{\lambda}=\sin^{2}\frac{\lambda\pi}{2\left(  N+1\right)  }
\label{eq:Delta1}%
\end{equation}
for $\lambda=1,2,\cdots,N$.
(An extra $\sigma_{x}$ pulse is required at $\tau_{N+1}=T$ for $N$ odd
\cite{West10}.)
It is convenient to consider the toggling frame associated with the qubit. In
this frame the qubit-environment Hamiltonian is modulated in time as:%
\begin{equation}
\tilde{H}\left(  \tau\right)  =F_{0}\left(  \tau/T\right)  S_{0}\otimes\hat
{B}_{0}\left(  \tau\right)  +F_{1}\left(  \tau/T\right)  S_{1}\otimes\hat
{B}_{1}\left(  \tau\right)  ,
\end{equation}
where $F_{\alpha}\left(  t\right)  =\left(  -1\right)  ^{\alpha\cdot\lambda}$
for $t\in\left(  \Delta_{\lambda},\Delta_{\lambda+1}\right]  $. The unitary
evolution operator of the toggling frame Hamiltonian is $\hat{U}\left(
T\right)  =\mathcal{T}\exp\left[  -i\int_{0}^{T}\tilde{H}\left(  \tau\right)
d\tau\right]  $, where $\mathcal{T}$ is the time-ordering operator. $\hat
{U}\left(  T\right)  $ has Dyson expansion%
\[
\sum_{s=0}^{\infty}\left(  -i\right)  ^{s}\sum_{\left\{  \alpha_{j}%
,p_{j}\right\}  }\hat{S}_{\left(  \oplus\alpha_{j}\right)  }\prod_{j=1}%
^{s}\hat{b}_{\alpha_{j},p_{j}}~\mathcal{F}_{\alpha_{1},\cdots,\alpha_{s}%
}^{p_{1},\cdots,p_{s}}~T^{s+p_{1}+\cdots p_{s}},
\]
where $\hat{S}_{\left(  \oplus\alpha_{j}\right)  }=\hat{S}_{\alpha_{1}}%
\cdots\hat{S}_{\alpha_{s}}$ and the coefficient $\mathcal{F}_{\alpha
_{1},\cdots,\alpha_{s}}^{p_{1},\cdots,p_{s}}$ can be obtained by the following
integral \cite{Yang08}
\begin{equation}
\mathcal{F}_{\alpha_{1},\cdots,\alpha_{s}}^{p_{1},\cdots,p_{s}}=\int_{0}%
^{1}dt_{s}\cdots\int_{0}^{t_{2}}dt_{1}\prod_{j=1}^{s}F_{\alpha_{j}}\left(
t_{j}\right)  t_{j}^{p_{j}}.
\end{equation}
When $\oplus_{j=1}^{s}\alpha_{j}=0$, the operator $\hat{S}_{\left(
\oplus\alpha_{j}\right)  }=I$ is the identity operator that acts trivially on
the qubit. Hence we only need to consider the terms with $\oplus_{j=1}%
^{s}\alpha_{j}\neq0$ that act non-trivially on the qubit. To show the
universality of the UDD sequence, we just need to prove $\hat{U}\left(
T\right)  =I\otimes\hat{U}_{B}\left(  T\right)  +O\left(  T^{N+1}\right)  $,
which can be reduced to verifying%
\begin{equation}
\mathcal{F}_{\alpha_{1},\cdots,\alpha_{s}}^{p_{1},\cdots,p_{s}}=0
\label{eq:FUDD}%
\end{equation}
for $\oplus_{j=1}^{s}\alpha_{j}\neq0$ and $s+\sum_{j=1}^{s}p_{j}\leq N$.
Eq.(\ref{eq:FUDD}) resembles the proof of Ref. \onlinecite{Yang08} for
universality of UDD. The key difference is that here additional indices
$\left\{  \alpha_{j}\right\}  $ are introduced to label different possible
qubit operators ($I$ and $\sigma_{z}$) which will be necessary for the proof
of universality of QDD and NUDD.

\paragraph*{QDD.---}

Let us now consider the QDD sequence \cite{West10} for generic interaction
between a single qubit and environment,%
\begin{equation}
H\left(  \tau\right)  =\sum_{\alpha}\hat{S}_{\alpha}\otimes\hat{B}_{\alpha
}\left(  \tau\right)  , \label{eq:H}%
\end{equation}
where $\hat{S}_{\alpha}=I,\sigma_{x},\sigma_{y},\sigma_{z}$ for binary vector
labels $\alpha=\left(  a_{2},a_{1}\right)  =\left(  0,0\right)  ,\left(
1,0\right)  ,\left(  1,1\right)  ,\left(  0,1\right)  $, respectively. For
Pauli matrices, one can verify that $\hat{S}_{\alpha}\hat{S}_{\alpha^{\prime}%
}=\pm\hat{S}_{\alpha\oplus\alpha^{\prime}}$, where $\oplus$ represents
pairwise binary addition without carry (e.g., $\left(  0,1\right)
\oplus\left(  0,1\right)  =\left(  0,0\right)  $). The QDD sequence consists
of two nesting levels of UDD with a total of $Q=\left(  N+1\right)  ^{2}$
pulses \cite{West10}.
The pulses $\sigma_{x}$ and $\sigma_{z}$ are associated with the first and
second levels, respectively. To label these $Q$ pulses, we introduce the
vector label $\lambda=\left(  l_{2},l_{1}\right)  \in\left\{  0,\cdots
,N\right\}  \otimes\left\{  1,\cdots,N+1\right\}  $ with $l_{2}\left(
N+1\right)  +l_{1}\in\left\{  1,\cdots,Q\right\}  $ \footnote{We use the Greek
letters $\alpha,\beta,\lambda,\kappa$ to represent vector labels, and use the
Latin letters $a_{r},b_{r},l_{r},k_{r}$ to represent the $r$-th element of the
corresponding vectors.}. The $\lambda$th pulse is applied at time
$\tau_{\lambda}=T\Delta_{\left(  l_{2},l_{1}\right)  }$ with%
\begin{equation}
\Delta_{\left(  l_{2},l_{1}\right)  }=\Delta_{l_{2}}+\left(  \Delta_{l_{2}%
+1}-\Delta_{l_{2}}\right)  \Delta_{l_{1}}.
\end{equation}
The toggling frame Hamiltonian for the QDD sequence is%
\begin{equation}
\tilde{H}\left(  \tau\right)  =\sum_{\alpha}F_{\alpha}\left(  \tau/T\right)
\hat{S}_{\alpha}\otimes\hat{B}_{\alpha}\left(  \tau\right)  \label{eq:Ht}%
\end{equation}
where $F_{\alpha}\left(  t\right)  =\left(  -1\right)  ^{a_{2}l_{2}+a_{1}%
l_{1}}$ for $t\in\left(  \Delta_{\left(  l_{2},l_{1}\right)  },\Delta_{\left(
l_{2},l_{1}+1\right)  }\right]  $. One can verify that
\begin{equation}
F_{\alpha}\left(  \tau/T\right)  F_{\alpha^{\prime}}\left(  \tau/T\right)
=F_{\alpha\oplus\alpha^{\prime}}\left(  \tau/T\right)  , \label{eq:Property4F}%
\end{equation}
which will be useful for our proof of universality of the QDD sequence. Using
the Dyson expansion of the unitary evolution operator of $\tilde{H}\left(
\tau\right)  $ for QDD, we obtain that to show the suppression of both the
dephasing and bit-flip errors up to $O\left(  T^{N+1}\right)  $ for small $T$,
it is sufficient to prove Eq.(\ref{eq:FUDD}) for $\oplus_{j=1}^{s}\alpha
_{j}\neq\left(  0,0\right)  $ and $s+\sum_{j=1}^{s}p_{j}\leq N$. This is very
similar to UDD, the difference being that $\alpha_{j}$ is now a two-component
binary vector.

\paragraph*{NUDD.---}

The NUDD sequence is a generalization of QDD from one-qubit to multiqubit
systems \cite{Mukhtar10a,Mukhtar10b,WangZY11a}. For $m$-qubit system, the most
general system-environment interaction can be written as Eq.(\ref{eq:H}) with
$\hat{S}_{\alpha}=\sigma_{v_{m}}^{\left(  m\right)  }\otimes\sigma_{v_{m-1}%
}^{\left(  m-1\right)  }\otimes\cdots\otimes\sigma_{v_{1}}^{\left(  1\right)
}$ and $\alpha=\left(  a_{2m},a_{2m-1},\cdots,a_{1}\right)  \in\left\{
0,1\right\}  ^{\otimes2m}$ for all generators. The Pauli operator of the $j$th
qubit is $\sigma_{v_{j}}^{\left(  j\right)  }=1,\sigma_{x}^{\left(  j\right)
},\sigma_{y}^{\left(  j\right)  },\sigma_{z}^{\left(  j\right)  }$ for
$\left(  a_{2j},a_{2j-1}\right)  =\left(  0,0\right)  ,\left(  1,0\right)
,\left(  1,1\right)  ,\left(  0,1\right)  $, respectively. The NUDD\ sequence
consists of $2m$ nesting levels and a total of $Q=\left(  N+1\right)  ^{2m}$
pulses. The decoupling pulse is $\sigma_{x}^{\left(  j\right)  }$ for $\left(
2j-1\right)  $th level and $\sigma_{z}^{\left(  j\right)  }$ for $2j$th level.
Similar to QDD, we introduce the label $\lambda=\left(  l_{2m},l_{2m-1}%
,\cdots,l_{1}\right)  \in\left\{  0,\cdots,N\right\}  ^{\otimes2m-1}%
\otimes\left\{  1,\cdots,N+1\right\}  $ with $\sum_{r=1}^{2m}l_{r}\left(
N+1\right)  ^{r-1}\in\left\{  1,\cdots,Q\right\}  $. The $\lambda$th pulse is
applied at time $\tau_{\lambda}=T\Delta_{\left(  l_{2m},\cdots,l_{1}\right)
}$, which is defined recursively%
\begin{equation}
\Delta_{\left(  l_{r},\cdots,l_{1}\right)  }=\Delta_{l_{r}}+\left(
\Delta_{l_{r}+1}-\Delta_{l_{r}}\right)  \Delta_{\left(  l_{r-1},\cdots
,l_{1}\right)  }\label{eq:Delta2}%
\end{equation}
for $r=2,\cdots,2m$. As illustrated in Fig.~\ref{fig:NUDD}, the timing for the
pulses has a self-similar structure \footnote{%
For $N$ odd, pulses at different levels may coincide and form one pulse. Hence
the unitary evolution of the $\lambda$th instantaneous pulse is:
$\tilde{\Omega}_{\left(  l_{2m},\cdots,l_{1}\neq N+1\right)  }=\Omega_{1}$,
$\tilde{\Omega}_{\left(  l_{2m},\cdots,l_{r}\neq N,N,\cdots,N,N+1\right)
}=\Omega_{r}\Omega_{r-1}^{N}\cdots\Omega_{1}^{N}$ for $r\geq2$, and
$\tilde{\Omega}_{\left(  N,\cdots,N+1\right)  }=\Omega_{2m}^{N}\cdots
\Omega_{1}^{N}$, where $\Omega_{2j-1}=\sigma_{x}^{\left(  j\right)  }$ and
$\Omega_{2j}=\sigma_{z}^{\left(  j\right)  }$.%
}. The toggling frame Hamiltonian for the NUDD\ sequence is the same as
Eq.(\ref{eq:Ht}), with $\alpha$ summed over all $4^{m}$ generators. Similar to
UDD and QDD, to show universality of the NUDD sequence, we just need to prove
Eq.(\ref{eq:FUDD}) for $\oplus_{j=1}^{s}\alpha_{j}\neq\vec{0}$ and
$s+\sum_{j=1}^{s}p_{j}\leq N$. We can view UDD\ and QDD\ as special cases of
NUDD with one and two nesting levels, respectively. Since the universality of
UDD, QDD, and NUDD all relies on verifying Eq.(\ref{eq:FUDD}), we will give a
general proof of Eq.(\ref{eq:FUDD}) in the rest of the paper.

\paragraph*{Universality Proof.---}

To prove Eq.(\ref{eq:FUDD}), we represent each integration over $t_{j}$ as a
linear operator acting on functions of $t_{j}$, which generates a function of
$t_{j+1}$. Thus, the process of multiple integrations can be thought of as a
discrete quantum walk in a functional space. We choose the basis of this
functional space according to the following consideration --- the functional
basis should be complete with respect to the operation $\int_{0}^{t}%
dt^{\prime}~F_{\alpha}\left(  t^{\prime}\right)  ~t^{p}$ for all relevant
orders up to $O\left(  T^{N+1}\right)  $. Since $F_{\alpha}\left(  t\right)  $
is a piece-wise analytic function with at most $Q=\left(  N+1\right)  ^{2m}$
discontinuities, it will be convenient to use piece-wise analytic functions as
our functional basis. In addition, we only need to consider all polynomials up
to power $N+1$ to characterize all the effects up to $O\left(  T^{N+1}\right)
$. Therefore, we choose basis that consists of $\left(  N+1\right)  \cdot Q$
functions%
\begin{equation}
\eta_{q,\lambda}\left(  t\right)  =t^{q}\eta_{\lambda}\left(  t\right)
\label{eq:eta}%
\end{equation}
with $\eta_{\lambda}\left(  t\right)  =$ $1$ for $t\in\left(  \Delta_{\lambda
},\Delta_{\lambda+1}\right]  $ and $\eta_{\lambda}\left(  t\right)  =0$
otherwise. Here $q\in\left\{  0,1,\cdots,N\right\}  $ is the \emph{polynomial
label} and $\lambda\in\left\{  0,\cdots,N\right\}  ^{\otimes2m}$ is the
\emph{pulse label}.

Then, we use Eq.(\ref{eq:Property4F}) to rewrite the integral%
\begin{align}
&  \mathcal{F}_{\alpha_{1},\cdots,\alpha_{s}}^{p_{1},\cdots,p_{s}}=\int
_{0}^{1}dt_{s}t_{s}^{p_{s}}\times\int_{0,\left[  \beta_{s-1}\right]  }^{t_{s}%
}dt_{s-1}t_{s-1}^{p_{s-1}}\nonumber\\
&  \cdots\int_{0,\left[  \beta_{1}\right]  }^{t_{2}}dt_{1}t_{1}^{p_{1}}\times
F_{\beta_{0}}\left(  t_{1}\right)  \times1, \label{eq:F2}%
\end{align}
where $\int_{0,\left[  \beta\right]  }^{t}dt^{\prime}\equiv F_{\beta}\left(
t\right)  \int_{0}^{t}dt^{\prime}F_{\beta}\left(  t^{\prime}\right)  $ and
$\beta_{j}=\oplus_{j^{\prime}=1}^{s-j}\alpha_{j^{\prime}}$. For $s+\sum
_{j=1}^{s}p_{j}\leq N$, we can compute the integral $\mathcal{F}_{\alpha
_{1},\cdots,\alpha_{s}}^{p_{1},\cdots,p_{s}}$ using functional basis $\left\{
\eta_{q,\lambda}\left(  t\right)  \right\}  $. For each operation $O$, let us
define the matrix form $\mathbf{O}_{q,\lambda}^{q^{\prime},\lambda^{\prime}}$
as $O\cdot\eta_{q,\lambda}=\mathbf{O}_{q,\lambda}^{q^{\prime},\lambda^{\prime
}}\eta_{q^{\prime},\lambda^{\prime}}$, where summation over repeating indices
is implied. For example, the multiplication $t\cdot\eta_{q,\lambda}%
=\eta_{q+1,\lambda}$ has the matrix form $\mathbf{M}_{q,\lambda}^{q^{\prime
},\lambda^{\prime}}=\delta_{q+1}^{q^{\prime}}\delta_{\lambda}^{\lambda
^{\prime}}$, with Kronecker delta function $\delta_{y}^{x}$. The other
operations are listed in Table~\ref{tab:1}.

Using block diagonal properties of matrices involved, $\mathcal{F}_{\alpha
_{1},\cdots,\alpha_{s}}^{p_{1},\cdots,p_{s}}$ can be further reduced as
multiplication of $Q\times Q$ sub-matrices $B_{\beta}$ and $D_{\beta}$
\cite{JI11b}
\begin{align}
&  \vec{v}_{L}\cdot\mathbf{M}^{p_{s}}\cdot\mathbf{G}_{\beta_{s-1}}%
\cdot\mathbf{M}^{p_{s-1}}\cdot\cdots\cdot\mathbf{G}_{\beta_{1}}\cdot
\mathbf{M}^{p_{1}}\cdot\mathbf{F}_{\beta_{0}}\vec{v}_{R}^{T}\nonumber\\
&  =\sum_{\left\{  i_{j}\geq0\right\}  }c_{i_{1},\cdots,i_{s}}^{\alpha
_{1},\alpha_{2},\cdots,\alpha_{s}}~\left\langle u_{L}\right\vert D_{\beta_{s}%
}^{i_{s}}D_{\beta_{s-1}}^{i_{s-1}}\cdots D_{\beta_{1}}^{i_{1}}B_{\beta_{0}%
}\left\vert u_{R}\right\rangle ,\nonumber
\end{align}
where $\sum_{j=1}^{s}i_{j}\leq s-1+\sum_{j=1}^{s}p_{j}\leq N-1$, $\beta
_{0}=\oplus_{j^{\prime}=1}^{s}\alpha_{j^{\prime}}$, and $c_{i_{1},\cdots
,i_{s}}^{\alpha_{1},\alpha_{2},\cdots,\alpha_{s}}$ are possibly non-zero
coefficients. The $Q$-vectors $\left\vert u_{L}\right\rangle $ and $\left\vert
u_{R}\right\rangle $ are determined by $\int_{0}^{1}dt_{s}$ and $1$ in
Eq.~(\ref{eq:F2}), respectively, with vector elements $\left\vert
u_{L}\right\rangle _{\lambda}=\Delta_{\lambda+1}-\Delta_{\lambda}$ and
$\left\vert u_{R}\right\rangle _{\lambda}=1$. The $Q\times Q$ matrices are%
\begin{align}
\left(  B_{\beta}\right)  _{\lambda}^{\lambda^{\prime}}  &  =\left(
-1\right)  ^{\beta\cdot\lambda}\delta_{\lambda}^{\lambda^{\prime}%
},\label{eq:BD}\\
\left(  D_{\beta}\right)  _{\lambda}^{\lambda^{\prime}}  &  =\Delta_{\lambda
}\delta_{\lambda}^{\lambda^{\prime}}-\left(  \Delta_{\lambda+1}-\Delta
_{\lambda}\right)  \sum_{\lambda^{\prime\prime}=\lambda+1}^{Q}\left(
-1\right)  ^{\beta\cdot\left(  \lambda^{\prime}-\lambda\right)  }%
\delta_{\lambda^{\prime\prime}}^{\lambda^{\prime}}.\nonumber
\end{align}
%

\begin{table}[b] \centering
\begin{tabular}
[c]{l|l}\hline
\textbf{Operation} & \textbf{Matrix/Vector Form}\\\hline\hline
$t\cdot\eta_{q,\lambda}=\eta_{q+1,\lambda}$ & $\mathbf{M}_{q,\lambda
}^{q^{\prime},\lambda^{\prime}}=\delta_{q+1}^{q^{\prime}}\delta_{\lambda
}^{\lambda^{\prime}}$\\\hline
$F_{\beta}\left(  t\right)  \cdot\eta_{q,\lambda}=\left(  -1\right)
^{\beta\cdot\lambda}\eta_{q,\lambda}$ & $\left(  \mathbf{F}_{\beta}\right)
_{q,\lambda}^{q^{\prime},\lambda^{\prime}}=\delta_{q}^{q^{\prime}}\left(
B_{\beta}\right)  _{\lambda}^{\lambda^{\prime}}$\\\hline
$\int_{0,\left[  \beta\right]  }^{t}dt^{\prime}\cdot\eta_{q,\lambda}=\left(
\mathbf{G}_{\beta}\right)  _{q,\lambda}^{q^{\prime},\lambda^{\prime}}%
\eta_{q^{\prime},\lambda^{\prime}}$ & $\left(  \mathbf{G}_{\beta}\right)
_{q,\lambda}^{q^{\prime},\lambda^{\prime}}=\frac{\delta_{q+1}^{q^{\prime}%
}\delta_{\lambda}^{\lambda^{\prime}}-\delta_{0}^{q^{\prime}}\left(  D_{\beta
}^{q+1}\right)  _{\lambda}^{\lambda^{\prime}}}{q+1}$\\\hline
$\int_{0}^{1}dt\cdot\eta_{q,\lambda}=\left(  \vec{v}_{L}\right)  _{q,\lambda}$
& $\left(  \vec{v}_{L}\right)  _{q,\lambda}=\frac{\Delta_{\lambda}%
^{q+1}-\Delta_{\lambda+1}^{q+1}}{q+1}$\\\hline
$1=\left(  \vec{v}_{R}^{T}\right)  _{q,\lambda}\eta_{q,\lambda}$ & $\left(
\vec{v}_{R}^{T}\right)  _{q,\lambda}=\delta_{0}^{q}$\\\hline
\end{tabular}
\caption{Matrix/vector forms of operations. Here $\delta_{y}^{x}$ is the Kronecker delta function, $B_{\beta}$ and
$D_{\beta}$ are $Q\times Q$ matrices as defined in Eq.~(\ref{eq:BD}).}\label{tab:1}%
\end{table}%

\paragraph*{Discrete Quantum Walk.---}

The last step is to verify
\begin{equation}
\left\langle u_{L}\right\vert D_{\beta_{s}}^{i_{s}}D_{\beta_{s-1}}^{i_{s-1}%
}\cdots D_{\beta_{1}}^{i_{1}}\left\vert B_{\beta_{0}}u_{R}\right\rangle =0
\label{eq:QW}%
\end{equation}
for $\beta_{0}\neq\vec{0}$ and $\sum_{j=1}^{s}i_{j}\leq N-1$. The left hand
side is the amplitude of a discrete \textquotedblleft quantum
walk\textquotedblright\ from initial state $\left\vert u_{L}\right\rangle $ to
a target state $\left\vert B_{\beta_{0}}u_{R}\right\rangle $\ in the
functional basis. Each multiplication of $D_{\beta}$ corresponds to one step
of a quantum walk. We need to show that the target state amplitude is
\emph{zero} when the number of steps is $\sum_{j=1}^{s}i_{j}\leq N-1$.%

\begin{figure*}%

\centering
\includegraphics[width=17cm]{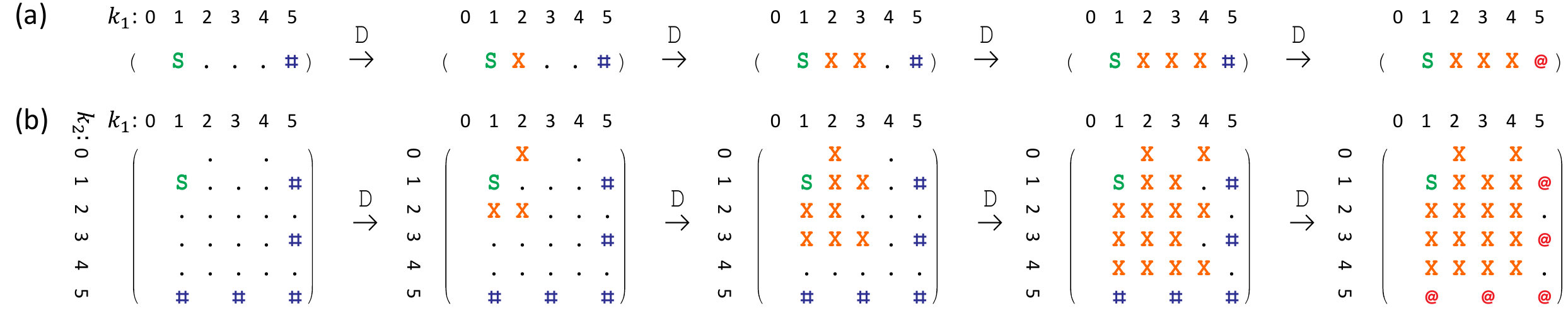}\caption[fig:Diffusion]{(color online). Evolution under discrete quantum walks for $N=4$. The symbols
represents different types of states: `\textcolor{PineGreen}{\textbf{S}}' for the initial
state, `\textcolor{orange}{\textbf{X}}' for states explored by a quantum walk,
`\textcolor{blue}{\textbf{\#}}' for unexplored target states, `\textcolor{red}{\textbf{@}}'
for explored target states, `\textcolor{black}{\textbf{.}}' for remaining unexplored
functional basis states. At least $N=4$ steps are necessary to reach the
target states, which is an illustration of Eq.~(\ref{eq:QW}). (a) For UDD, a
discrete quantum walk occurs in a one-dimensional functional basis, with
initial state $\left\vert 1\right\rangle $, target state $\left\vert
N+1\right\rangle $, and quantum walk matrix $D_{1}$. (b) For QDD, a discrete
quantum walk occurs in a two-dimensional functional basis, with initial state
$\left\vert 11\right\rangle $, target states $\left\{  \left\vert \kappa
^{\#}\right\rangle \right\}  $, and a quantum walk matrix $D_{\left(
01\right)  }$.}\label{fig:Diffusion}
%

\end{figure*}%

To understand the quantum walk we choose a convenient basis
\begin{equation}
\chi_{\kappa}\left(  t\right)  =\sum_{\lambda}c_{\kappa,\lambda}\eta_{\lambda
}\left(  t\right)
\end{equation}
with orthogonal transformation%
\begin{equation}
c_{\kappa,\lambda}=\prod_{r=1}^{2m}\sin\left[  \frac{\left(  2l_{r}+1\right)
k_{r}+\left(  N+1\right)  \left(  k_{r-1}-1\right)  }{N+1}\frac{\pi}%
{2}\right]  ,
\end{equation}
where $\kappa=\left(  k_{2m},k_{2m-1},\cdots,k_{1}\right)  \in\left\{
0,\cdots,N+1\right\}  ^{\otimes2m}$ and $k_{0}=1$ \footnote{Note that exactly
$\left(  N+1\right)  ^{2m}$ of $\left(  N+2\right)  ^{2m}$ possible
$\chi_{\kappa}$'s are non-zero functions, which form a new complete orthogonal
basis. Other functions are simply zeros (e.g., $\chi_{\left(  k_{2m}%
,\cdots,k_{2},0\right)  }=\chi_{\left(  k_{2m},\cdots,k_{r+2},N+1,2h_{r}%
,\cdots\right)  }=\chi_{\left(  k_{2m},\cdots,k_{r+2},0,2h_{r}+1,\cdots
\right)  }=0$), which are kept for notational convenience.}. In the basis
$\left\{  \left\vert \kappa\right\rangle \right\}  $ with $\left\vert
\kappa\right\rangle \equiv\chi_{\kappa}$, the initial state is $\left\vert
u_{L}\right\rangle \propto\left\vert \left(  1,\cdots,1\right)  \right\rangle
$. Since $\beta_{0}\neq\vec{0}$ (with $b_{r^{\ast}}=1$ and $b_{r<r^{\ast}}%
=0$), the target state is $\left\vert B_{\beta_{0}}u_{R}\right\rangle
=\sum_{\kappa^{\#}}u_{\beta_{0},\kappa^{\#}}\left\vert \kappa^{\#}%
\right\rangle $ where the $r^{\ast}$th element of $\kappa^{\#}$ is
$k_{r^{\ast}}^{\#}=N+1$. After some calculation \cite{JI11b}, it can be shown
that $D_{\beta}\left\vert \kappa\right\rangle =\sum_{k_{1}^{\prime}=0}%
^{k_{1}+1}\cdots\sum_{k_{2m}^{\prime}=0}^{k_{2m}+1}d_{\kappa^{\prime},\kappa
}\left\vert \kappa^{\prime}\right\rangle $ where $d_{\kappa^{\prime},\kappa}$
denotes possibly non-zero coefficients. Since each step of quantum walk only
increases the index $k_{r}$ by at most $1$ unit, it requires at least $N$
steps to walk from $k_{r^{\ast}}=1$ to $k_{r^{\ast}}^{\#}=N+1$. Therefore,
there is zero amplitude in the target states when the number of steps
$\sum_{j=1}^{s}i_{j}\leq N-1$. This completes the proof of Eq.(\ref{eq:QW})
and implies the universality of NUDD. We emphasize that after the basis in
Eq.~(\ref{eq:eta}) is introduced, all operations are matrix multiplications,
and all our analytical statements have been explicitly checked numerically.

We can illustrate the quantum walk for special cases: (1) For UDD, the
orthogonal transformation $c_{\kappa,\lambda}=c_{\left(  k_{1}\right)
,\left(  l_{1}\right)  }=\sin\left[  k_{1}\frac{2l_{1}+1}{N+1}\frac{\pi}%
{2}\right]  $ is simply the Fourier transformation as in
Ref.~\onlinecite{Yang08}. The functional basis $\left\{  \left\vert
\kappa\right\rangle \right\}  $ with $\kappa=\left(  k_{1}\right)  \in\left\{
1,\cdots,N+1\right\}  $ forms a one-dimensional array. As illustrated in
Fig.~\ref{fig:Diffusion}a, it will take at least $N$ steps to walk from
$\kappa=\left(  1\right)  $ to $\kappa^{\#}=\left(  N+1\right)  $. (2) For
QDD, the orthogonal transformation is a little more complicated $c_{\kappa
,\lambda}=c_{\left(  k_{2},k_{1}\right)  ,\left(  l_{2},l_{1}\right)
}=\left(  -1\right)  ^{(k_{1}-1)/2}\sin\left[  k_{2}\frac{2l_{2}+1}{N+1}%
\frac{\pi}{2}\right]  \sin\left[  k_{1}\frac{2l_{1}+1}{N+1}\frac{\pi}%
{2}\right]  $ for odd $k_{1}$ and $\left(  -1\right)  ^{k_{1}/2}\cos\left[
k_{2}\frac{2l_{2}+1}{N+1}\frac{\pi}{2}\right]  \sin\left[  k_{1}\frac
{2l_{1}+1}{N+1}\frac{\pi}{2}\right]  $ for even $k_{1}$. The functional basis
$\left\{  \left\vert \kappa\right\rangle \right\}  $ with $\kappa=\left(
k_{2},k_{1}\right)  $ forms a two-dimensional array. As illustrated in
Fig.~\ref{fig:Diffusion}b, it will also take at least $N$ steps to walk from
$\kappa=\left(  1,1\right)  $ to $\kappa^{\#}=\left(  N+1,k_{1}\right)  $ or
$\left(  k_{2},N+1\right)  $. (3) For NUDD, the functional basis forms a
$2m$-dimensional array. Similar to UDD and QDD, it will take at least $N$
steps to walk from $\kappa=\left(  1,\cdots,1\right)  $ to $\kappa^{\#}$ with
$k_{r}^{\#}=N+1$.


When the suppression order is $N_{r}$ for the $r$th nesting level of NUDD, the
overall suppression of decoherence is $O\left(  T^{N^{\ast}+1}\right)  $
limited by the lowest suppression order $N^{\ast}=\min\left[  N_{1}%
,\cdots,N_{2m}\right]  $.

\paragraph*{Summary \& Outlook.---}

We have proved the universality of the NUDD sequence, which can restore an
unknown initial state of $m$-qubit system to $O\left(  T^{N+1}\right)  $ using
$2m$ nesting levels and $\left(  N+1\right)  ^{2m}$ pulses, independent of the
details of the system-environment interaction. The NUDD\ sequence is
experimentally feasible, because it requires only $poly\left(  N\right)  $
pulses acting on individual qubits. Our work illustrates a general connection
between DD problems and discrete quantum walks. The techniques developed can
be used to address a variety of questions, such as investigation of
environment correlations using DD, schemes of efficient DD for particular
system-environment interaction, and the combination of multiqubit DD schemes
with quantum error correcting codes \cite{Ng09} and quantum algorithms.

\paragraph*{Note added.---}
After submission of this work, a preprint of closely related work by Kuo and Lidar \cite{Kuo11} became available, with a different proof of the universality of QDD.


We would like to thank John Preskill for helpful discussions. This work was
supported by the Sherman Fairchild Foundation,  A.P. Sloan Foundation under Grant No.
BR-5123, and NSF Career Award DMR-1049082


\end{document}